\documentclass[journal]{IEEEtran}
\usepackage{amsmath,amsfonts}
\usepackage{algorithmic}
\usepackage{algorithm}

\usepackage{array}

\usepackage[caption=false,font=normalsize,labelfont=sf,textfont=sf]{subfig}
\usepackage{textcomp}
\usepackage{stfloats}
\usepackage{float}
\usepackage{url}
\usepackage{verbatim}
\usepackage[justification=centering]{caption}
\usepackage{graphicx}
\usepackage{cite}
\newtheorem{strategy}{Strategy}
\usepackage{amssymb}
\usepackage{hyperref}
\usepackage{multirow}
\usepackage{xcolor}
\newtheorem{definition}{Definition}

\usepackage[caption=false]{subfig}

\begin{document}

\title{Targeted Mining of Time-Interval Related Patterns}

\author{Shuang Liang,~\IEEEmembership{Member,~IEEE,} Lili Chen, Wensheng Gan,~\IEEEmembership{Member,~IEEE}, \\ Philip S. Yu,~\IEEEmembership{Life Fellow,~IEEE}, Shengjie Zhao,~\IEEEmembership{Member,~IEEE}

\thanks{This work was supported in part by the National Natural Science Foundation of China (No. 62076183), Shanghai Science and Technology Innovation Action Project (No. 20511100700), Shanghai Science and Technology Commission Project (No. 23511103100), Shanghai Municipal Science and Technology Major Project (No. 2021SHZDZX0100), and the Fundamental Research Funds for the Central Universities. (Corresponding author: Shuang Liang)
}
    
\thanks{Shuang Liang, Lili Chen, and Shengjie Zhao are with the School of Computer Science and Technology, Tongji University, Shanghai 201804, China. (E-mail: shuangliang@tongji.edu.cn, lilichien3@gmail.com, shengjiezhao@tongji.edu.cn)}

\thanks{Wensheng Gan is with the College of Cyber Security, Jinan University, Guangzhou 510632, China. (E-mail: wsgan001@gmail.com)}
	
\thanks{Philip S. Yu is with the Department of Computer Science,  University of Illinois Chicago, Chicago, USA. (E-mail: psyu@uic.edu)} 

}

\markboth{IEEE Trans. XXX~2025}%
{Shell \MakeLowercase{\textit{et al.}}: A Sample Article Using IEEEtran.cls for IEEE Journals}

\maketitle

\begin{abstract}
    Compared to frequent pattern mining, sequential pattern mining emphasizes the temporal aspect and finds broad applications across various fields. However, numerous studies treat temporal events as single time points, neglecting their durations. Time-interval-related pattern (TIRP) mining is introduced to address this issue and has been applied to healthcare analytics, stock prediction, etc. Typically, mining all patterns is not only computationally challenging for accurate forecasting but also resource-intensive in terms of time and memory. Targeting the extraction of time-interval-related patterns based on specific criteria can improve data analysis efficiency and better align with customer preferences. Therefore, this paper proposes a novel algorithm called TaTIRP to discover \textbf{T}argeted  \textbf{T}ime-\textbf{I}nterval  \textbf{R}elated  \textbf{P}attern. Additionally, we develop multiple pruning strategies to eliminate redundant extension operations, thereby enhancing performance on large-scale datasets. Finally, we conduct experiments on various real-world and synthetic datasets to validate the accuracy and efficiency of the proposed algorithm.
  
\end{abstract}


\section{Introduction}
\IEEEPARstart{W}{ith} the rapid advancements in information technology and the widespread adoption of sensor devices, vast amounts of raw data are being generated. Data mining enables the exploration and analysis of this information, allowing valuable knowledge to be extracted. Frequent pattern mining (FPM) \cite{agrawal1994fast} was introduced to identify patterns that frequently appear in transaction databases, where their occurrences exceed a threshold specified by an expert. In FPM, all items in a transaction are unordered, and a consensus is that they occur simultaneously. Nevertheless, behavioral data are collected over time \cite{gan2020fast} and space \cite{bao2021mining, yang2022mining}, and sequential data with temporal order is a more appropriate description of human life. Sequential pattern mining (SPM) \cite{agrawal1995mining, fournier2017survey, gan2020proum, sharmiladevi2021closed} considers $\langle A, B \rangle$  and $\langle B, A \rangle$ as two distinct sequential patterns. For instance, if a patient visits a hospital and reports a cough followed by a fever, this scenario is diagnosed differently from a fever followed by a cough, and the treatment would vary accordingly.  Numerous algorithms have been proposed to efficiently extract frequently occurring sequential patterns from sequence databases. SPM has been widely applied in numerous fields such as anomaly detection \cite{boniol2020automated}, autonomous driving \cite{teng2023motion}, and recommendation systems \cite{roy2022systematic}.

\begin{figure}
    \centering
    \includegraphics[trim=0 0 0 0,clip,scale=0.35]{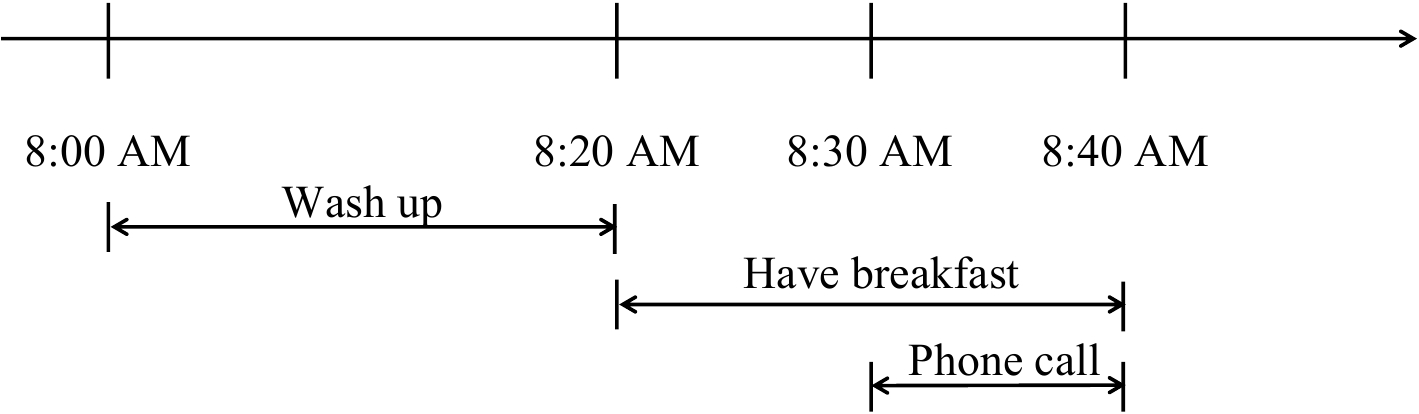}
    \caption{Sequence of events after waking up}
    \label{fig:intervalsequence}
\end{figure}

The above studies represent each event with a timestamp and deem that all events occur either simultaneously or sequentially, neglecting the fact that there is a required duration for the event to take place. For example, as shown in Fig. \ref{fig:intervalsequence}, a person washed up from 8:00 AM to 8:20 AM and had breakfast from 8:20 AM to 8:40 AM. During breakfast, the person also took a 10-minute phone call. If we adopt the timestamp approach, it would be impossible to indicate whether the phone call occurred during or after breakfast. To tackle this challenge, the time-interval-related pattern (TIRP) \cite{lee2020z, moskovitch2015classification, mordvanyuk2021verttirp, fournier2022fasttirp, wu2007mining, papapetrou2009mining} has been suggested for a richer representation, which includes duration, order of occurrence, and temporal relationship of events.  Frequent TIRP mining aims to identify patterns within a time-interval sequence database that exhibit support exceeding a specified threshold. In comparison with sequential pattern mining, TIRP mining is more sophisticated, as it computes event durations and determines temporal relationships between events, such as when event  $A$ intersects with or occurs before event $B$.

There have been several studies to address the TIRP mining task. Initially, the temporal relations between different events are simply assumed to be a complementary process for the dual events \cite{kam2000discovering}, indicating that event $A$ occurring before event $B$ is equivalent to event $B$ occurring after event $A$. The disadvantage of this method is that it is hard to determine the temporal relation between the first and third events when more than two events occur. For instance, $A$ is before $B$ and $B$ is before $C$. Easily, we have that $A$ is before $C$. Nevertheless, suppose $A$ occurs before $B$ and $C$ also occurs before $B$, then the relation between $A$ and $C$ is inaccessible. Fortunately, TPrefixSpan \cite{wu2007mining} is developed to discover non-ambiguous temporal patterns. Afterward, Papapetrou \cite{papapetrou2009mining} proposed a noise tolerance method to address the ambiguity in Allen's temporal relations \cite{allen1994actions}. In this context, we may assume that meeting at 10:00 is the same as meeting at 10:02. After that, KarmaLego \cite{moskovitch2015classification} employs the epsilon margin and leverages the transitivity property inherent in Allen’s temporal relations. Subsequently, VertTIRP \cite{mordvanyuk2021verttirp} is introduced to form the unambiguous definition of temporal relations and utilize a vertical representation that harnesses the transitivity property of these temporal relations. Recently, FastTIRP \cite{fournier2022fasttirp} implemented a Pair Support Pruning (PSP) method to reduce the frequency of join operations.

These algorithms become highly inefficient when processing extensive databases, making it difficult to derive accurate forecasts from the obtained results. This inefficiency stems from the generation of a vast number of non-valuable sequential patterns throughout the discovery operation.  To capture the attention of customers, targeted query \cite{huang2023taspm, hu2024targeted, miao2022targeted, zhang2022tusq, gan2022towards} emerges as a tailored exploitation technique. For example, a targeted pattern query is dedicated to discovering all related frequent patterns when clients put pencils and books on their shopping lists. A few approaches have been proposed for targeted queries, such as targeted sequential pattern querying \cite{huang2023taspm}, targeted mining of contiguous sequential patterns \cite{hu2024targeted}, targeted high-utility itemset querying \cite{miao2022targeted}, and targeted high-utility sequence querying \cite{zhang2022tusq}. To the best of our knowledge, no research has been conducted on TIRP mining using targeted queries. Implementing targeted queries while simultaneously maximizing efficiency in the context of time-interval-related pattern mining is challenging, despite its wide range of application scenarios. For example, through personalized recommendations, we can gather customer intentions and analyze continuous monitoring data from patients (such as blood glucose, blood pressure, and electrocardiogram) to identify TIRPs in disease progression. These patterns help in developing personalized treatment plans and preventive measures. By understanding the timing and progression of a patient's condition, healthcare providers can make more informed decisions regarding medication adjustments, lifestyle recommendations, and early interventions, ultimately improving patient outcomes and reducing healthcare costs. In the financial market, by analyzing subscriber preferences and high-frequency data (such as trading volume and price movement), we can identify TIRPs to optimize high-frequency trading algorithms and enhance trading returns. These patterns enable algorithmic trading systems to better predict market trends based on historical data, allowing real-time, data-driven decisions that maximize profit potential and minimize risk. Similarly, by analyzing user activity data across different time-intervals, we can uncover user behavior patterns and provide personalized content recommendations and targeted advertisements, improving marketing effectiveness. 

To tackle the aforementioned issues, this paper introduces a novel mining task, termed targeted mining of time-interval-related patterns, and presents an efficient algorithm called TaTIRP. The primary contributions are outlined as follows:

\begin{itemize}
    \item This paper introduces a novel mining task—targeted time-interval-related pattern (TIRP) mining. Unlike traditional sequential pattern mining methods, TIRP not only takes into account the order of events but also incorporates the time-intervals and durations between events, offering a more accurate and enriched representation of patterns. Additionally, this paper innovatively combines targeted queries with TIRP mining, enabling the precise extraction of patterns that align with specific objectives based on user needs. By reducing the generation of irrelevant patterns, targeted queries significantly improve mining efficiency, making the process more efficient and better tailored to the user's preferences.
 
    \item To enhance the efficiency, we introduce three pruning strategies: USFP, which eliminates sequences that do not contain the query event; UEPP, which ceases expansion when the support between the pattern and extension event fails to satisfy the required criteria; and UQPP, which halts expansion if the support between the query and extension event does not meet the specified conditions. Furthermore, we leverage the Pair Support Matrix (PSM) to assess the support of event pairs.
    	
    \item Subsequent experiments conducted on real and synthetic datasets demonstrate that our proposed TaTIRP achieves superior performance. Furthermore, most unpromising patterns are effectively pruned during the excavation process using the proposed pruning strategies.
\end{itemize}

The remainder of this paper is structured as follows. Section \ref{Related work} provides a comprehensive review of related literature. Section \ref{Preliminary and problem statement} elucidates the preliminaries and problem statement. Section \ref{Algorithms} introduces the designed pruning strategies and the algorithm formulated based on these strategies. Section \ref{Experiment} delineates the experimental findings and their analysis. Section \ref{Conclusion} concludes the paper and proposes avenues for future research.

\section{Related work}  \label{Related work}
\subsection{Sequential pattern mining}

The sequential pattern mining framework was first established by Agrawal and Srikant, who also introduced the AprioriAll algorithm \cite{agrawal1995mining}, the pioneering algorithm specifically designed for SPM. This algorithm is based on the Apriori property, which asserts that all subsequences of a frequent sequence are also frequent. This property significantly reduces the number of candidate patterns, thereby enhancing the efficiency of the mining process. Following this, the generalized sequential pattern (GSP) algorithm \cite{srikant1996mining} was introduced to further improve the efficiency and applicability of sequential pattern mining. The GSP algorithm enhances the capability to handle complex sequence data by iteratively generating candidate sequences and detecting frequent patterns. Subsequently, Zaki et al. \cite{zaki2001spade} presented the sequential pattern discovery using equivalent classes (SPADE) algorithm, which stores sequences in a vertical data format and utilizes equivalence class partitioning and intersection calculations to discover frequent sequences, significantly enhancing computational efficiency. Meanwhile,  the prefix-projected sequential pattern mining (PrefixSpan) algorithm \cite{han2001prefixspan} utilizes prefix projection for sequence pattern mining, thereby avoiding the generation of candidate sequences and further optimizing the mining process. Later, the sequential pattern mining (SPAM) algorithm \cite{ayres2002sequential} is based on a vertical data format and utilizes bitmap operations to extract sequential patterns. This approach effectively enhances the capability to handle large-scale datasets. The last pattern (LAPIN) algorithm \cite{yang2007lapin} can optimize the mining of long sequential patterns, thereby improving performance when handling extended sequences. Li et al. \cite{salvemini2011fast} suggested frequent pattern mining using a sequential testing (FAST) algorithm, which employs sequential testing for frequent pattern mining, optimizing both the time complexity and space complexity of this algorithm. Additionally, Fournier-Viger et al. \cite{fournier2014fast} have further enhanced the efficiency of sequential pattern mining by integrating compression techniques with classical algorithms. The co-occurrence Map is employed to achieve an order-of-magnitude improvement over other algorithms on several datasets. Beyond the conventional SPM algorithms previously mentioned, several other variants have been introduced, including non-overlapping SPM \cite{wu2017nosep, wu2020netncsp}, maximal SPM \cite{fournier2014vmsp, li2022netnmsp}, closed SPM \cite{yan2003clospan,fumarola2016clofast}, and high-utility SPM \cite{gan2021explainable, gan2021utility}. Overall, the development of the sequential pattern mining field has consistently focused on improving algorithm efficiency, expanding the applicability of these algorithms, and optimizing the ability to handle large-scale datasets. With the continuous advancements in big data and artificial intelligence technologies, sequential pattern mining is poised to play an increasingly vital role in business decision-making, public services, and scientific research. Advancements in sequential pattern mining have significantly enhanced both efficiency and applicability. However, specialized patterns, such as time-interval related patterns, require more sophisticated methods. The following subsection delves into how time-interval related pattern mining extends and refines the capabilities of sequential pattern mining.

\subsection{Targeted pattern querying}

Sequential pattern mining is essential for extracting meaningful patterns from sequence databases, offering powerful analytical capabilities across diverse fields. Nevertheless, the substantial quantity of frequent sub-sequences produced by SPM presents a considerable obstacle to efficient pattern analysis. To address this, the thought of targeted query has arisen, catering to scenarios where specific events or sequences are of primary interest to decision-makers. Targeted query tasks focus on discovering patterns that align with user-defined query sequences, thereby enhancing the relevance and efficiency of pattern extraction. This approach accelerates the mining process by reducing the exploration of irrelevant patterns, thereby streamlining data analysis tasks. Over time, targeted query techniques have evolved into a crucial area of research and practical application, finding utility in fields like bioinformatics and confidentiality management. Early efforts in targeted query were pioneered in frequent pattern mining, where Kubat et al. \cite{kubat2003itemset} introduced the Itemset Tree (IT) structure for querying and investigating transaction data. However, IT structures often require substantial memory resources. In response, Fournier-Viger et al. \cite{fournier2013meit} proposed the Memory Efficient Itemset Tree (MEIT), which optimizes memory usage through compression mechanisms within the IT structure. For scenarios involving multiple items, the GFP-growth algorithm is developed by Shabtay et al. \cite{shabtay2021guided}, which proficiently balances time and memory efficiency. Similarly, Miao et al. \cite{miao2022towards} introduced the THUIM algorithm, integrating targeted queries into high-utility pattern mining via a list-centric pattern comparison approach. In the realm of sequence data, Chand et al. \cite{chand2012target} addressed financial limitations in the context of targeted sequential pattern mining tasks, requiring the query item to be the last in a pattern sequence. Recently, Huang et al. \cite{huang2023taspm} suggested the TaSPM algorithm, depicting the dataset through bitmap representation and leveraging co-occurrence pruning techniques to enhance efficiency in generalized targeted SPM tasks. In addition, Zhang et al. \cite{zhang2022tusq} adopted a projection technology utilizing the targeted chain and developed suffix remains utility and terminated descendants utility upper bounds to obtain excellent performance in targeted utility-oriented sequence querying issues. In summary, advancements in targeted queries continue refining techniques for efficient and effective pattern discovery in sequence databases. These innovations address a wide range of analytical challenges across diverse real-world applications, paving the way for enhanced data-driven decision-making and insights. However, research on targeted queries in TIRP remains limited, presenting an opportunity for further innovation, particularly in improving querying efficiency. While TaSPM \cite{huang2023taspm} enhances the efficiency of sequence pattern mining by optimizing target query sequences, it does not specifically address time-interval patterns. In contrast, our algorithm not only innovates in sequence pattern mining but also boosts the relevance of target mining by considering event durations and temporal relationships. Our algorithm aims to bridge this gap by exploring novel approaches that integrate targeted queries into the TIRP framework, improving both scalability and performance, while also implementing pruning strategies to further optimize efficiency.

\subsection{Time-interval related pattern mining}

Initially, Kam and Fu \cite{kam2000discovering} introduced the TIRP mining algorithm but highlighted challenges such as unknown temporal relations between events and the variability in expressing the same TIRP, leading to inconsistent pattern counts. Wu et al. \cite{wu2007mining} addressed the ambiguity in TIRP representation by proposing a canonical approach. This method standardizes the representation of temporal relationships between events by sorting them based on their temporal attributes, aiming to enhance pattern mining consistency, and the TPrefixSpan algorithm \cite{wu2007mining} was adapted specifically for TIRP mining built upon the canonical representation concept. This approach efficiently discovers temporal patterns by leveraging the sorted event sequences. Thereafter, IEMiner \cite{patel2008mining} extended Kam's work by incorporating a vector-based representation of temporal relation counts and further adopting the canonical approach. However, challenges remained in managing temporal ambiguities, particularly with larger TIRPs. KarmaLego \cite{moskovitch2009medical} was introduced with a matrix-based representation to tackle temporal relation ambiguities. This method stored relations between event pairs in a sorted manner based on their temporal characteristics, thereby improving the efficiency of pattern discovery. DharmaLego \cite{moskovitch2015outcomes} is a refinement of KarmaLego that utilizes a hash table to store instances of smaller TIRPs. This enhancement aimed to accelerate the mining process, especially in handling large-scale datasets. Subsequently, vertTIRP \cite{mordvanyuk2021verttirp} formulated a robust definition of temporal relations that reduces ambiguity, particularly when considering uncertainties in event start and end times, and enhances efficiency by utilizing a pairing strategy. FastTIRP \cite{fournier2022fasttirp} incorporated a novel Pair Support Pruning (PSP) technique to streamline the exploration of potential patterns, thereby reducing computational complexity. Recent advancements have also explored mining top-k high-utility patterns \cite{huang2019mining}, similarity matching on time-interval sequences \cite{mirbagheri2020similarity}, and sub-sequence search of event sequences \cite{yang2017subsequence}. These approaches contribute to diverse mining tasks beyond traditional frequency-based pattern discovery. Compared to existing time-interval pattern mining methods, our algorithm, built upon the latest FastTIRP \cite{fournier2022fasttirp}, significantly enhances both efficiency and relevance. In terms of efficiency, the three pruning strategies (USFP, UQPP, and UEPP) effectively reduce unnecessary computations, particularly for large datasets and complex queries. By precisely matching target query event sequences, these strategies improve runtime efficiency. USFP eliminates redundant calculations early on, while UQPP and UEPP optimize extensions from the query and extend event perspectives, reducing memory usage and processing time in big data scenarios. In terms of relevance, our algorithm focuses on accurately mining target query event sequences rather than broad frequent patterns. This approach allows it to uncover time-interval patterns directly related to the query, improving pattern quality, especially when dealing with complex event durations, overlaps, and containment. Overall, our algorithm provides a significant boost in both efficiency and relevance, offering a more effective solution for real-world applications.

\section{Preliminary and problem statement}
\label{Preliminary and problem statement}

In this section, we first introduce the fundamental definitions of time-interval-related patterns and targeted querying, with the principal notation summarized in Table \ref{table:notation}. We then describe the problem statement of time-interval-related pattern mining with targeted constraints.

\begin{table}[!htbp]
    \centering
    \footnotesize
    \caption{Notations}
    \label{table:notation}
	\begin{tabular}{|c|l|}
		\hline
		\textbf{Notation} & \textbf{Explanation}  \\ \hline
		$ e $ & 	event types \\ \hline
         $A$ &	such as $A$, $B$, $C$, \ldots,  represent events \\ \hline
		$ E $ &   a collection of events \\ \hline
        $ \epsilon $ &	  noise parameter \\ \hline
        $t$ &	  time point \\ \hline
         =$_{\epsilon} $ &	  quasi-equal \\ \hline
         $\prec_{\epsilon}$ &	  precedes by at least $\epsilon$ \\ \hline
         $ I $ &	   symbolic time-interval, a triplet $I$ = (\textit{start}, \textit{end}, $e$) \\ \hline
         $S$ &	time-interval sequence  \\ \hline
          $\prec$ &	  the ordering of symbolic time-intervals \\ \hline
          $\mathcal{D}$ &	time-interval sequence database  \\ \hline
         \textit{minGap}  &	  the minimum gap between two symbolic time-intervals\\ \hline
        \textit{maxGap}  &	the maximum gap between two symbolic time-intervals \\ \hline
        \textit{minDura}  &	the minimum duration of a symbolic time-intervals\\ \hline
        \textit{maxDura} &	 the maximum duration of a symbolic time-intervals \\ \hline
        $\Phi$ &	 the set of all event types  in the running example \\ \hline
        $\Gamma$ &	 all relations  in the running example \\ \hline
        $R_{ij}$ &	 the relation between the corresponding $i$-th and $j$-th event \\ \hline
        $R$ &	 the set of relations in $E$ \\ \hline
         
        TIRP &   time-interval related pattern, $(E, R)$ \\ \hline
        S-TIRP &   a set aggregating multiple TIRPs that share the same events  \\ \hline
         \textit{VSup} &	vertical support \\ \hline
         \textit{HSup} &	horizontal support \\ \hline
          \textit{minSup} &	a specified minimum support \\ \hline
          \textit{qes} &	 query event sequence  \\ \hline
          $\widehat{\textit{ts}}$ &	 target S-TIRP  \\ \hline         
	\end{tabular}
\end{table}

\subsection{Basic preliminaries}

\begin{definition}
  \rm Assuming there exists a collection of event types $E$ = \{$e_1$, $e_2$, $\dots$, $e_n$\} that is organized in lexicographical order. A symbolic time-interval $I$ is characterized as a triplet $I$ = (\textit{start}, \textit{end}, $e$). In this context, \textit{start} indicates the beginning time, and \textit{end} denotes the ending time of an event $e$.
\end{definition}

For brevity, the terms \textit{I.start}, \textit{I.end}, and \textit{I.e} are assigned, respectively, to represent the values in the above triple. Furthermore, instead of adopting an exact operator to compare two time-intervals, we are prone to employing an approximate ($\epsilon$) method to deal with noise \cite{moskovitch2015classification}. 

\begin{figure}[!htbp]
    \centering
    \includegraphics[trim=0 0 0 0,clip,scale=0.43]{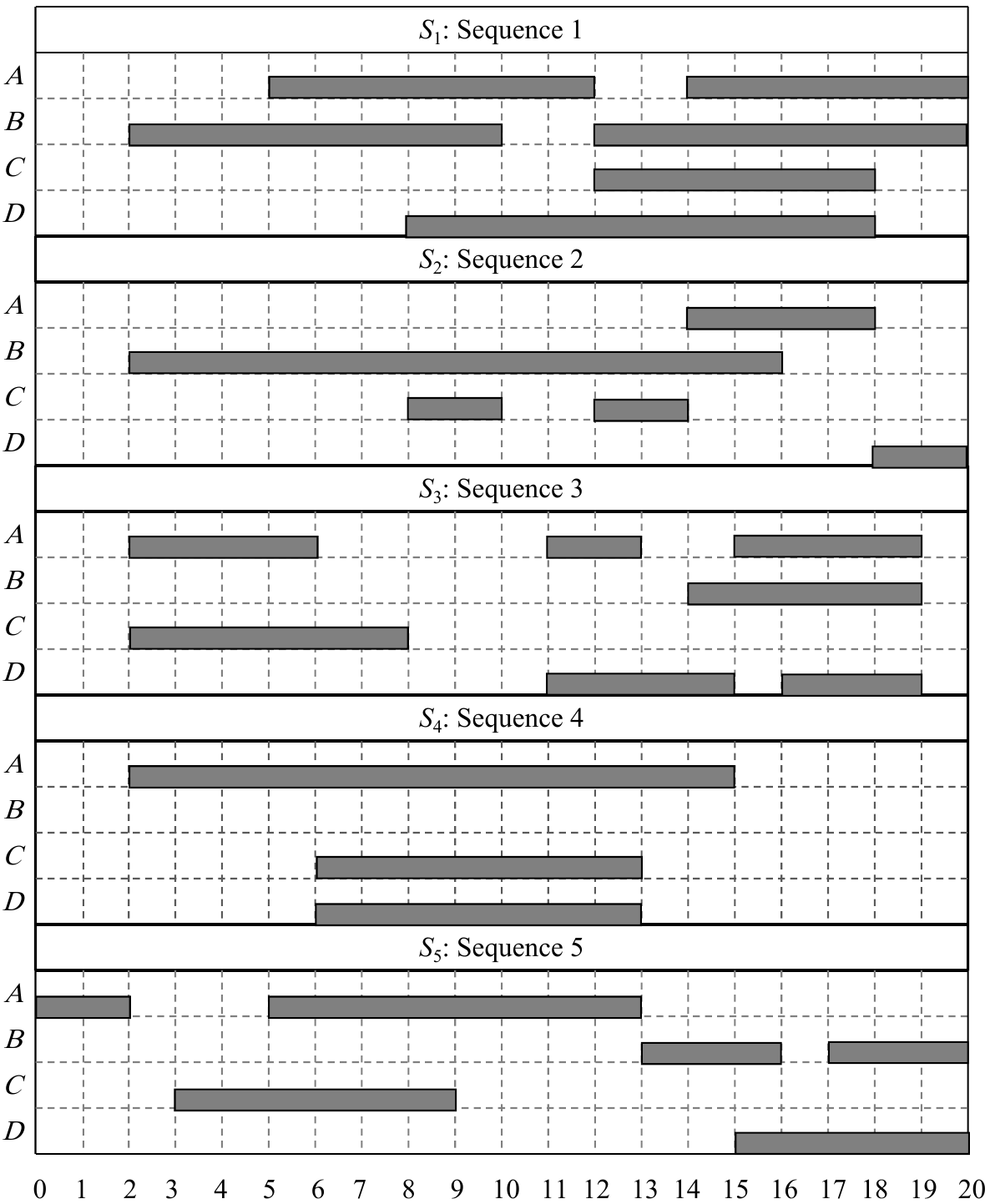}
    \caption{ A time-interval sequence database.}
    \label{fig:example database}
\end{figure}

\begin{definition}
   \rm Suppose there exists a given noise number $\epsilon$ $\geq 0$, and two time points $t_i$ and $t_j$ satisfy $|t_i$ - $t_j|$ $\leq$ $\epsilon$, then the two time points are quasi-equal and expressed as $t_i$ =$_{\epsilon}$       $t_j$. Additionally, if $(t_j$ - $t_i)$ $>$ $\epsilon$, it can be expressed as $t_i$ $\prec_{\epsilon}$ $t_j$. These notations allow us to arrange time-interval events in a sequence.
\end{definition}

\begin{definition}
   \label{def_prec}
   \rm Let a finite time-interval sequence \textit{S} = \{\textit{I}$_{1}$, \textit{I}$_{2}$, $\ldots$, \textit{I$_{m}$}\}  consist of $m$ distinct symbolic time-intervals following the order of $\prec$. In the case of $I_i \prec I_j$, we have ($I_i.\textit{start} \prec_{\epsilon} I_j.\textit{start}$) $\vee$ ($I_i.\textit{start} =_{\epsilon} I_j.\textit{start} \wedge I_i.\textit{end} \prec_{\epsilon} I_j.\textit{end}$) $\vee$ 
    ($I_i.\textit{start} =_{\epsilon} I_j.\textit{start} \wedge I_i.\textit{end} =_{\epsilon} I_j.\textit{end} \wedge I_i.e \prec I_j.e$).
\end{definition}

\begin{definition}
    \rm A time-interval sequence database $\mathcal{D}$ is composed of a collection of time-interval sequences $\mathcal{D}$ = $\langle S_1, S_2, \ldots, S_n \rangle$. Each sequence $S_i$ is assigned a unique identifier $i$.
\end{definition}

For example, as illustrated in Fig. \ref{fig:example database}, the time-interval sequence database is composed of five sequences. Note that two identical events in a sequence cannot occur at the same time point. In the sequence $S_1$, the event $A$ starts at time 5 and ends at time 12, then starts at time 14 and ends at time 20; the event $B$ starts at time 2 and ends at time 10, then starts at time 12 and ends at time 20; the event $C$ starts at time 12 and ends at time 18, while the event $D$ starts at time 8 and ends at time 18.  According to the rules of sorting prescribed in Definition \ref{def_prec}, $S_1$ can be represented as \{(2, 10, \textit{B}), (5, 12, \textit{A}), (8, 18, \textit{D}), (12, 18, \textit{C}), (12, 20, \textit{B}), (14, 20, \textit{A})\}. Similarly, the original database is transformed into a horizontal database $\mathcal{HD}$ in $\prec$ order as shown in Table \ref{table:db}.

\begin{table*}[!htbp]
	\centering
	\caption{A horizontal database}
	\label{table:db}
	\begin{tabular}{|c|c|}
		\hline
		\textbf{\textit{SID}} & \textbf{Time-interval sequence}  \\ \hline
		$ S_{1} $ & 	(2, 10, \textit{B}), (5, 12, \textit{A}), (8, 18, \textit{D}), (12, 18, \textit{C}), (12, 20, \textit{B}), (14, 20, \textit{A})\\ \hline
		$ S_{2} $ & (2, 16, \textit{B}), (8, 10, \textit{C}), (12, 14, \textit{C}), (14, 18, \textit{A}), (18, 20, \textit{D}) \\ \hline
		$ S_{3} $ &	  (2, 6, \textit{A}), (2, 8, \textit{C}), (11, 13, \textit{A}), (11, 15, \textit{D}), (14, 19, \textit{B}), (15, 19, \textit{A}), (16, 19, \textit{D})  \\ \hline
        $ S_{4} $ &	  (2, 15, \textit{A}), (6, 13, \textit{C}), (6, 13, \textit{D})
   \\ \hline
        $ S_{5} $ &	  (0, 2, \textit{A}), (3, 9, \textit{C}), (5, 13, \textit{A}), (13, 16, \textit{B}), (15, 20, \textit{D}), (17, 20, \textit{B})
   \\ \hline
	\end{tabular}
\end{table*}

\begin{definition}
   \rm  We define the temporal relationship between any pair of symbolic time-intervals $I_i$ and $I_j$ as $r(I_i, I_j)$. In aggregate, there are eight temporal relations, which are \textit{before} ($b$), \textit{meet} ($m$), \textit{overlap} ($o$), \textit{contain} ($c$), \textit{finished by} ($f$), \textit{equal} ($e$), \textit{start} ($s$), \textit{left contain} ($l$).
\end{definition}

\begin{figure*}[!htbp]
    \centering
    \includegraphics[trim=0 0 0 0,clip,scale=0.48]{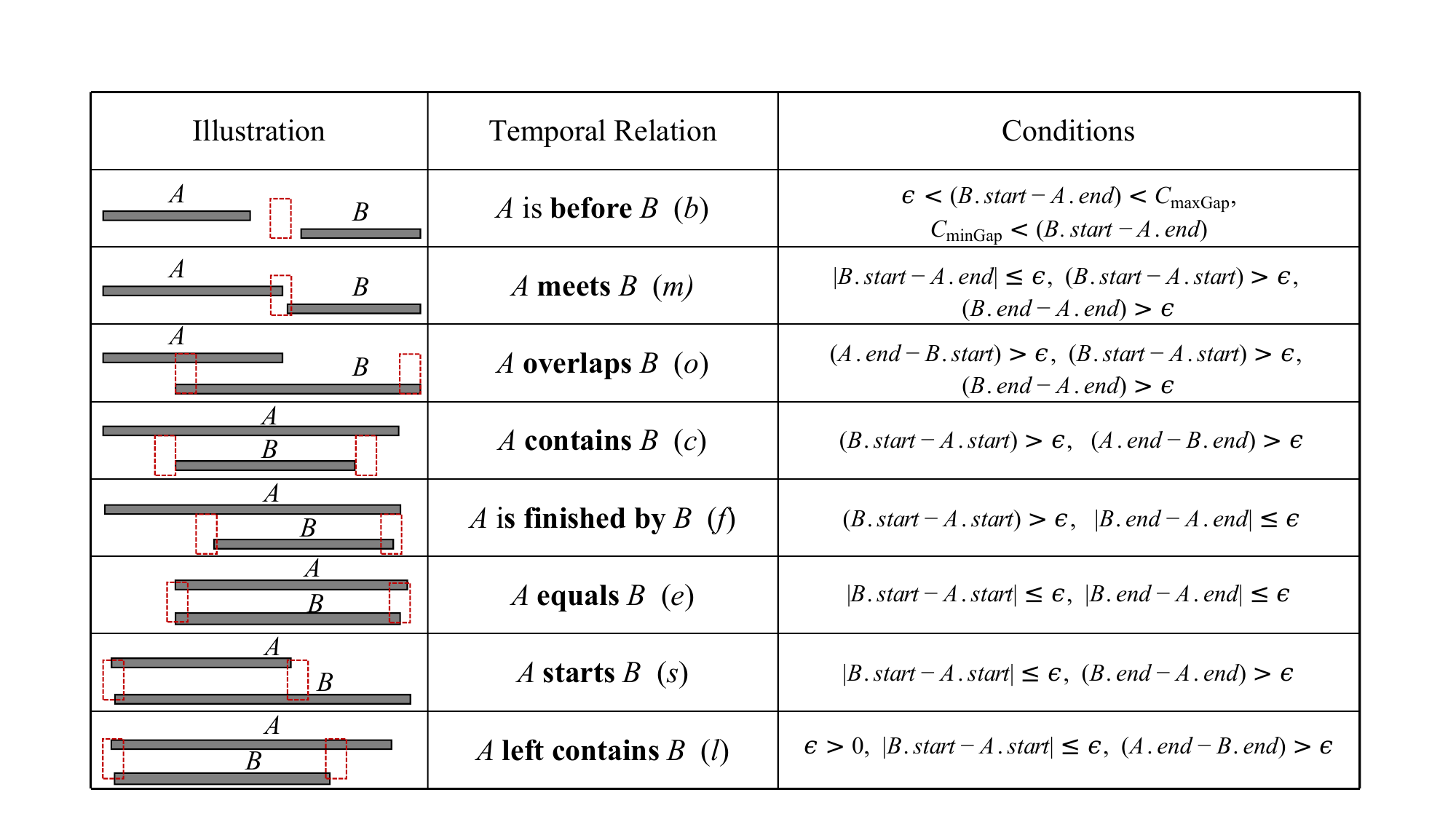}
    \caption{Relations and the corresponding conditions.}
    \label{fig:relation}
\end{figure*}

Several alternative constraints, \textit{minGap}, \textit{maxGap}, \textit{minDura}, and \textit{maxDura} are used to describe these relations better. Among them, \textit{minGap} and \textit{maxGap} are used to confine the gap between the two symbolic time-intervals when their relations are defined as "before". On the other hand, \textit{minDura} and \textit{maxDura} require the time-interval of a pattern to be within a certain range.  The conditions to be fulfilled for each temporal relation and corresponding illustrations are exhibited in Fig. \ref{fig:relation}. The red dashed box in the figure indicates the noise scale. Since the duration of all patterns is required to be in a certain range, they are not specifically listed therein.  In these relations, it is crucial to distinguish between \textit{start} and \textit{left contain} to avoid ambiguity. The primary distinction is that in the \textit{start} relation, $B$.\textit{end} exceeds $A$.\textit{end}, whereas in the \textit{left contain} relation, $B$.\textit{end} is smaller than $A$.\textit{end}.

\begin{definition}
    \rm  Assuming that the set of all event types is $\Phi$ = $\{A, B, C, D\}$ in the running example and that of all relations is $\Gamma$ = $\{b, m, o, c, f, e, s, l\}$, $E$ and $R$ are ordered subsets of $\Phi$ and $\Gamma$, respectively. A time-interval related pattern (TIRP) is denoted as $X$ = $(E, R)$ where $R$ = \{$r(I_1, I_2), r(I_1, I_3), \ldots, r(I_2, I_3), \ldots, r(I_{k-1}, I_k)$\} and $R_{ij}$ is the relation between the corresponding $i$-th and $j$-th event in $E$.
\end{definition}

For example, assuming that \textit{minGap} is set to 0, \textit{maxGap} is 5, \textit{minDura} is 0, and \textit{maxDura} is 20. The TIRP $(\textit{BA}, o)$ appears in the sequence $S_1$ with \{(2, 10, \textit{B}), (5, 12, \textit{A})\}, $(\textit{BA}, b)$ exists in $S_1$ with \{(2, 10, \textit{B}), (14, 20, \textit{A})\}.  Besides, the TIRP $(\textit{BAC}, obm)$ occurs in the sequence $S_1$ with \{(2, 10, \textit{B}), (5, 12, \textit{A}), (12, 18, \textit{C})\}. According to the \textit{maxGap} constraint, $(\textit{CB}, b)$ does not appear in $S_3$, because the gap (\textit{B.end }- \textit{C.start}) is larger than 5. In the running example, we can observe that even though the two events are the same, their relationships may vary in the same or different sequences such that the temporal relations between events $B$ and $A$ are \textit{overlap} and \textit{before} in the sequence $S_1$.

\begin{definition}
    \rm  We aggregate multiple time-interval related patterns sharing the same events $X$ into a set called S-TIRP, which is denoted as $\widehat{X}$ = $(X, \_)$. Supposing that there exists a TIRP $X$ = $(E, R)$ and a time-interval sequence $S$ = \{\textit{I}$_{1}$, \textit{I}$_{2}$, $\ldots$, \textit{I$_{m}$}\}, we regard $X$ matches $S$ when the former is contained within the latter. In other words, for any events $e_i$ and $e_j$ in $E$, there exist $I_k$ and $I_h$ in $S$ satisfying that $e_i$ = $I_k.e$, $e_j$ = $I_h.e$ and $r(I_k, I_h)$ = $R_{ij}$.  
\end{definition}

Matching TIRPs with time-interval sequences needs consideration of both events and relations, which can be exceedingly stringent in practical mining processes. To uncover more interesting patterns, we propose a broader definition of matching.

\begin{definition}
   \rm  Let there be an S-TIRP $\widehat{X}$ = ($E, \_$) with  $E$ = \{$e_1$, $e_2$, $\ldots$, $e_s$\} and a time-interval sequence $S$ = \{\textit{I}$_{1}$, \textit{I}$_{2}$, $\ldots$, \textit{I$_{m}$}\}, we regard $\widehat{X}$ matches $S$ if $\forall$ 1 $\leq v \leq s$,  $\exists k_v$ such that 1 $\leq k_1 < k_2 < \ldots < k_v \leq m$, $e_v$ = \textit{I}$_{k_v}.e$.  
\end{definition}

As mentioned in the above example, the TIRP $(\textit{BA}, o)$ matches $S_1$ and $S_2$. In addition, the S-TIRP $\widehat{\textit{BA}}$ = $\{(\textit{BA}, o), (\textit{BA}, b), (\textit{BA}, f)\}$ matches $S_1$ and $S_2$. 

\begin{definition}
   \rm The vertical support (\textit{VSup}) of an S-TIRP $\widehat{X}$ is defined as the number of sequences matching $\widehat{X}$. If $\textit{VSup}(\widehat{X})$ exceeds the parameter $ \textit{minSup} \times |\mathcal{D}|$ where \textit{minSup} is a percentage of the database size established by the expert and $|\mathcal{D}|$ is the size of the given database, then $\widehat{X}$ is a frequent S-TIRP. The horizontal support (\textit{HSup}) of $\widehat{X}$ is defined as the number of times that $\widehat{X}$ matches a time-interval sequence $S$. 
\end{definition}

Continuing with the example, the vertical support of  $\widehat{\textit{BA}}$ is  \textit{VSup($\widehat{\textit{BA}}$)} = 2. Besides, $\widehat{\textit{BA}}$ is found 3 times in sequence $S_1$, thus \textit{HSup($\widehat{\textit{BA}}$, $S_1$)} = 3.

In contrast to the horizontal database, a vertical database proposed by \cite{Zaki2001} of an S-TIRP has been proven to be a more functional representation, which is a table containing multiple attributes of S-TIRP.

\begin{figure}
    \centering
    \includegraphics[trim=1 0 0 0,clip,scale=0.42]{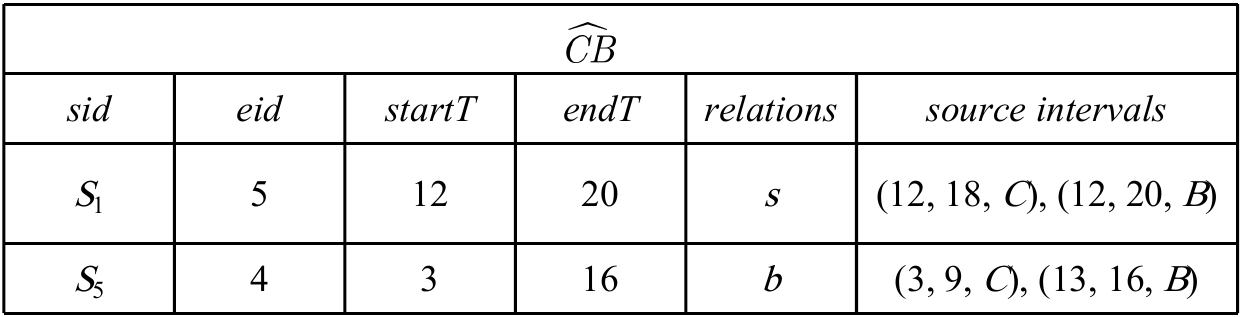}
    \caption{ The vertical database of $\widehat{\textit{CB}}$.}
    \label{fig:table}
\end{figure}

\begin{definition}
   \rm  A vertical database of an S-TIRP $\widehat{X}$ includes a list of sequence IDs and extends with the event ID, the start and end time, the source time-intervals, and respective relations in the SID sequence. For convenience of description, an illustrative vertical database is depicted in Fig. \ref{fig:table}.
\end{definition}

In the sequence $S_1$, \textit{eid} represents the position of the last symbolic time-interval of $\widehat{\textit{CB}}$, which is 5. The \textit{startT} and \textit{endT} are the minimum of the start time and the maximum of the end time of $\widehat{\textit{C}}$ and $\widehat{\textit{B}}$, respectively. According to the conditions of relations, their relation is denoted as $s$. The source interval records symbolic time-intervals comprising $\widehat{\textit{CB}}$.

To be more consistent with customer requirements, it is meaningful to explore targeted patterns with query patterns from users rather than excavating a large number of patterns.

\begin{definition}
  \rm  Given a query event sequence \textit{qes} = \{$e_1$, $e_2$, $\ldots$, $e_k$\},  the target S-TIRP $\widehat{\textit{ts}}$ = ($E, \_$) of \textit{qes} will adhere to  \textit{qes} $\subseteq$ $E$ and \textit{VSup}($\widehat{\textit{ts}}$) $\geq$ \textit{minSup} $\times$ $|\mathcal{D}|$ where $|\mathcal{D}|$ represents the size of the given database. Targeted TIRP mining aims to uncover a collection of target S-TIRPs with \textit{qes}.
\end{definition}

\subsection{Problem statement}

Considering a time-interval sequence database $\mathcal{D}$, a query event sequence \textit{qes}, and a specified minimum support \textit{minSup}, and some optional constraints such as \textit{minGap}, \textit{maxGap}, \textit{minDura}, \textit{maxDura}, $\epsilon$, the objective of the targeted time-interval related pattern is to identify the complete set of target S-TIRPs $\widehat{\textit{ts}}$  = ($E, \_$) that meet the following criteria: \textit{qes} $\subseteq$ $E$, \textit{VSup($\widehat{\textit{ts}}$)} $\geq$ \textit{minSup} $\times$ $|\mathcal{D}|$.

For instance, in Table \ref{table:db}, concerning a particular query event sequence \textit{qes} = \{$A$, $C$\} and the prescribed \textit{minSup} is configured as 0.4. The size of the given time-interval sequence database is 5. Besides, \textit{minGap} is set to 0, \textit{maxGap} is 5, \textit{minDura} is 0, \textit{maxDura} is 20, and $\epsilon$ is 0. Based on the above conditions, there are a total of eight S-TIRPs exploited, which are $\widehat{\textit{AC}}$, $\widehat{\textit{ACA}}$, $\widehat{\textit{ACAB}}$, $\widehat{\textit{ACABD}}$, $\widehat{\textit{ACAD}}$, $\widehat{\textit{ACADB}}$, $\widehat{\textit{ACB}}$, $\widehat{\textit{ACD}}$.

Assuming $A$, $B$, $C$, and $D$ represent different online user behaviors: $A$ for user login, $B$ for browsing products, $C$ for viewing product details, and $D$ for completing a purchase. In this context, the mined patterns highlight frequently occurring activity sequences that include  \textit{AC}. , where \textit{A} happens first, followed by \textit{C}. For instance, the pattern  $\widehat{\textit{ACADB}}$ suggests that the user logs in, views product details, logs in again to continue browsing products, and ultimately completes the purchase. This pattern may indicate that after repeatedly viewing product details, the user decides to make a purchase, reflecting their decision-making process.

\section{Algorithms}  \label{Algorithms}

In this section, we design several pruning strategies to mitigate memory consumption and time expenditure. Furthermore, we present the proposed algorithm accompanied by its pseudocode.

\subsection{Proposed pruning strategies}

Typically, there are two kinds of sequential pattern growth, namely $S$-expansion and $I$-expansion. The former is employed when a new event happens at a different moment, and the latter is utilized when multiple events are accomplished simultaneously. 

\begin{definition}
    \rm  $S$-expansion, referred to as sequence extension, generates an event sequence by appending an event following the final event of the sequence pattern $X$, which is denoted as  $X'$ = $X \bigotimes e$.
\end{definition}

\begin{definition}
    \rm  $I$-expansion, known as itemset extension, describes the process of appending an event to the last event of the sequence pattern $X$, which is symbolized as  $X'$ = $X \bigoplus e$.
\end{definition}

In the current context, no identical symbolic time-interval exists due to the defined ordering rules that strictly order each symbolic time-interval. Therefore, we only extend the pattern by $S$-expansion in this paper.

\begin{definition}
   \rm  The event corresponding to the matching position in the query event sequence is termed the current query event, represented as \textit{qe}. Throughout the pattern growth process, we employ a flag \textit{match} to track the current match position of \textit{qe}. When an extended event $e$ does not match \textit{qe}, the \textit{match} remains 0. Conversely, upon events $e$ and \textit{qe} being identical, \textit{match} is updated by adding 1.  If the \textit{match} attains the length of the query event sequence, it signifies that the sequence has been entirely matched.
\end{definition}

Since the events of target S-TIRPs necessarily contain the query event sequence \textit{qes}, we advocate for the adoption of the subsequent pruning strategy based on this condition.

\begin{strategy}
    \label{stra_1}
    (Unpromising Sequence Filter Pruning Strategy, USFP): For a query event sequence \textit{qes}, any time-interval sequence $S$ in the database $\mathcal{D}$ that does not match $\widehat{\textit{qes}}$ is discarded. Sequences lacking $\widehat{\textit{qes}}$ cannot contribute to generating the target S-TIRPs. By eliminating these non-contributing sequences, the TaTIRP approach not only mitigates memory consumption but also optimizes processing performance.  Suppose the size of the database $\mathcal{D}$ falls below the minimum support threshold \textit{minSup} $\times$ $|\mathcal{D}|$, it indicates that no frequent target S-TIRP can be discovered.
\end{strategy}

In the running example, the query event sequence \textit{qes} = \{$A$, $C$\}, we can observe that only $S_2$ does not contain \textit{qes}; thus, we can remove this sequence directly from the database, which will substantially decrease the search space for pattern growth. In the case where the targeted query pattern is absent from the dataset, Strategy \ref{stra_1} will filter out all sequences that do not contain the query event sequence. As a result, no further processing will be required, as no valid patterns can be found in the remaining sequences. FastTIRP \cite{fournier2022fasttirp} adopts a data structure called Pattern Support Matrix by storing the co-occurrence between two S-TIRPs to minimize the number of join operations required. Here, we continue to proceed along this structure to optimize the algorithm.

\begin{definition}
   \rm For each pair of event types $e_1$ and $e_2$ $\in E$ in a time-interval sequence database $\mathcal{D}$, where $e_1$ is temporally before $e_2$, the Pair Support Matrix (PSM) maintains a triple where PSM($e_1$, $e_2$) =  \textit{VSup($\widehat{e_1e_2}$)}.
\end{definition}

\begin{figure}[!htbp]
    \centering
    \includegraphics[trim=0 0 0 0,clip,scale=0.43]{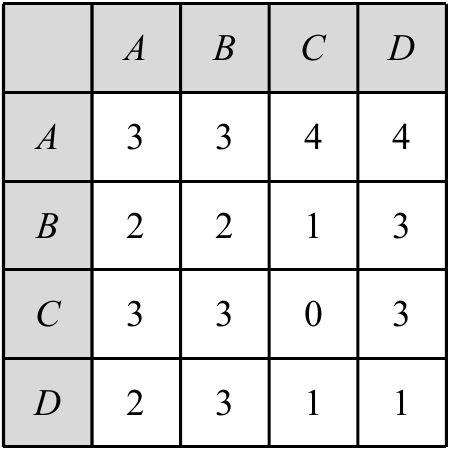}
    \caption{ The PSM structure.}
    \label{fig:PSM}
\end{figure}

Fig. \ref{fig:PSM} illustrates the construction of the PSM structure where the sequence $S_2$ has been removed. When calculating PSM($A$, $B$), it is equivalent to \textit{VSup}($\widehat{AB}$). In the database, sequence $S_1$ contains TIRP $(AB, m)$, where event $A$  occurs between time 5 and 12, and event $B$ takes place from time 12 to 20. Sequence $S_3$ contains TIRP $(AB, o)$, with event $A$ occurring from time 11 to 13, followed by event $B$ from time 14 to 19. Sequence $S_5$ comprises TIRP $(AB, m)$, where event $A$ occurs between time 5 and 13, and event $B$ transpires from time 13 to 16. Moreover, sequence $S_5$ also incorporates TIRP $(AB, b)$, where event $A$ spans from time 5 to 13, and event $B$ occurs between time 17 and 20. In conclusion, as $\widehat{AB}$ appears in three sequences, the value of PSM($A$, $B$) is 3. Similarly, the other values are calculated in the same way. By leveraging the structure, two pruning strategies are designed, with Strategy \ref{stra_2} having already been proposed in the FastTIRP algorithm \cite{fournier2022fasttirp}.

\begin{strategy}
    \label{stra_3}
    (Unpromising Query Pattern Pruning Strategy, UQPP): If the support of the last event $p$ in the S-TIRP $\widehat{X}$, when combined with the current query event $qe$, falls below the threshold (e.g., PSM($p$, $qe$) $<$ \textit{minSup} $\times$ $|\mathcal{D}|$), then $\widehat{X}$ cannot contribute to the growth pattern. This is because the support of the TIRP \textit{VSup}($\widehat{Xqe}$) is lower than PSM($p$, $qe$), resulting in \textit{VSup}($\widehat{Xqe}$) $<$ \textit{minSup} $\times$ $|\mathcal{D}|$).
\end{strategy}

Strategy \ref{stra_3} focuses on the co-occurrence and temporal relationships between the current pattern and the query events, while the following suggested Strategy \ref{stra_2} highlights these relationships between the current pattern and the extension events.

\begin{strategy}
    \label{stra_2}
  (Unpromising Extension Pattern Pruning Strategy, UEPP): If the support of the last event, $p$, of the S-TIRP  with an event $q$ to be extended does not surpass the support threshold, e.g., PSM($p$, $q$) $<$ \textit{minSup} $\times |\mathcal{D}|$, it is superfluous to incorporate $q$ into the pattern to construct a more comprehensive pattern. Due to the downward closure property of support, no extension of an S-TIRP $\widehat{\textit{pq}}$ is frequent \cite{fournier2022fasttirp}.
\end{strategy}

While constructing the PSM structure, FastTIRP \cite{fournier2022fasttirp} requires that the duration of two S-TIRPs is less than \textit{maxDura} and the gap of them is less than \textit{maxGap}. In actual implementation, we identify that considering the gap limit in the initial phase prunes out patterns that could have been frequent. For example, there are four symbolic time-intervals in a sequence: (0, 20, \textit{A}), (0, 3, \textit{B}), (5, 8, \textit{C}), and (15, 20, \textit{C}). We can detect that the gap of the S-TIRP $\widehat{\textit{CC}}$ is equal to 7, which is greater than \textit{maxGap}. Nevertheless, $\widehat{\textit{ABCC}}$ is eligible in the final output. This phenomenon is inconsistent with the downward closure property of support. This is because the start time and end time of $\widehat{\textit{AB}}$ are 0 and 20, and that of $\widehat{\textit{ABC}}$ are also 0 and 20. The extendable conditions are satisfied between $\widehat{\textit{ABC}}$ and $\widehat{\textit{C}}$. Consequently, $\widehat{\textit{ABCC}}$ is sensible. Hence, in light of the above example, we eliminate the gap restriction during the construction of the PSM structure.

Although both Strategy \ref{stra_3} and Strategy \ref{stra_2} are based on the PSM structure, a distinction still exists. Strategy \ref{stra_2} focuses on the extended event, while Strategy \ref{stra_3} is centered on the current query event. For example, the query event sequence \textit{qes} = \{$C$, $B$\} and \textit{minSup} is set as 0.4. According to Strategy \ref{stra_1}, since the sequence $S_2$ does not contain \textit{qes}, $S_2$ can be directly removed.  Furthermore, assuming that the current event is $D$ and the event to be extended is $C$, according to PSM($D$, $C$) $ < $ \textit{minSup} $\times$ $|\mathcal{D}|$ (Strategy \ref{stra_2}), the extension can be stopped. Suppose the last event of the current pattern is $B$ and the current query event \textit{qe} is $C$, there is no need to extend the pattern due to PSM($B$, $C$) $ < $ \textit{minSup} $\times$ $|\mathcal{D}|$ (Strategy \ref{stra_3}).

\subsection{Proposed TaTIRP algorithm}

The following describes the concrete procedure of TaTIRP and the application of several pruning strategies proposed above. Firstly, the pseudocode of the formulated TaTIRP approach is demonstrated in Algorithm \ref{TaTirp}.

\begin{algorithm}[h]
    \small
    \caption{The TaTIRP algorithm}
    \begin{algorithmic}[1]	
    \label{TaTirp}
    \REQUIRE a time-interval sequence database $\mathcal{D}$, a query event sequence \textit{qes} provided by the user, the prescribed minimum support threshold, \textit{minSup}, optional parameters $\epsilon$, \textit{minGap}, \textit{maxGap}, \textit{minDura}, \textit{maxDura}.
    \ENSURE complete targeted S-TIRPs of \textit{qes}.

	 \STATE Scan the time-interval sequence database $\mathcal{D}$  to create horizontal database $\mathcal{HD}$;
	\STATE Filter sequences that do not contain \textit{qes} to derive \textit{HD'}; (\textbf{USFP Strategy})
	\STATE Scan the $\mathcal{HD'}$ to obtain all frequent S-TIRP stored in \textit{SF} and construct a vertical database $\mathcal{VD}$  for each S-TIRP sorting in descending order of frequency. 
	\STATE Construct PSM according to $\mathcal{HD'}$;
        \STATE \textit{match} = 0;
		\FOR {$\widehat{P} \in$ \textit{SF}}
    		 \STATE \textbf{call} \textbf{DepthFirstSearch}($\widehat{P}$, \textit{qes}, \textit{SF}, \textit{minSup}, \textit{match});
		\ENDFOR
\end{algorithmic}
\end{algorithm}

The input of the algorithm consists of a time-interval sequence database $\mathcal{D}$,  a query event sequence \textit{qes} provided by users, the required minimum support \textit{minSup}, and some optional parameters such as noise margin $\epsilon$, the minimum gap between two time points \textit{minGap}, the maximum gap between two time-points \textit{maxGap}, the minimum duration between two S-TIRPs \textit{minDura}, and the maximum duration between two S-TIRPs \textit{maxDura}.  The output of the algorithm is all targeted S-TIRPs of \textit{qes}.  TaTIRP first scans the time-interval sequence database $\mathcal{D}$ to create a horizontal database $\mathcal{HD}$. The horizontal database includes multiple sequences composed of a collection of symbolic time-intervals sorted in a defined order $\prec$.  Then, filter sequences that do not contain \textit{qes} to derive $\mathcal{HD'}$ by utilizing the USFP strategy. Sequences not containing \textit{qes} cannot be supporting sequences for the target pattern. Afterward, scan the updated horizontal database $\mathcal{HD'}$ to obtain all the frequent S-TIRPs and store them in a set \textit{SF}. Then, construct a vertical database $\mathcal{VD}$ for each frequent S-TIRP and sort in descending order of frequency.  In the vertical database, it is necessary to record the last position of the events $eid$ appearing in a sequence $S_{sid}$, the start time, the end time, the temporal relations, and the symbolic time-intervals incorporated. The crucial information can be utilized directly for pattern extension, avoiding multiple scans of the sequence database. Next, construct the PSM structure according to $\mathcal{HD'}$, preparing for the following pruning strategies. Since we need to recognize the position of the input \textit{qes} matches, the variable \textit{match} is introduced to record the matching position between the query event sequence and the current pattern. For each S-TIRP $\widehat{P}$ in \textit{SF}, to acquire a longer S-TIRP, the TaTIRP will perform the DepthFirstSearch process to implement pattern growth.

\begin{algorithm}[h]
    \small
    \caption{DepthFirstSearch}
    \begin{algorithmic}[1]		
    \label{DepthFirstSearch}
    \REQUIRE a prefix S-TIRP $\widehat{P}$, a query event sequence \textit{qes}, a set containing all frequent S-TIRPs \textit{SF} that can perform extension, the required minimum support \textit{minSup}, the matching position of the query event sequence \textit{match}.
    \ENSURE all target S-TIRPs of \textit{qes}.
	
	\STATE \textit{lastEventOfP} = the last event of $\widehat{P}$;
        \STATE \textit{qe} = the current query event;
        
        \IF{\textit{lastEventOfP} == \textit{qe}}
			 \STATE \textit{match}++;
		\ENDIF
  
       \IF{\textit{match} == \textit{qes}.size}
			\STATE Output($\widehat{P}$);
		\ENDIF
  
		\IF {PSM(\textit{lastEventOfP}, \textit{qe}) $<$ \textit{minSup} $\times |\mathcal{D}|$}
		     \STATE return null; (\textbf{UQPP Strategy})
		\ENDIF
  
		\FOR {$\widehat{F} \in$ \textit{SF}}
		\STATE \textit{new\_match} = \textit{match};
		\IF {PSM(\textit{lastEventOfP}, $F$)  $<$ \textit{minSup} $\times |\mathcal{D}|$}
		\STATE continue; (\textbf{UEPP Strategy})
		\ENDIF
		\STATE Construct a vertical database for $\widehat{\textit{PF}}$;
		\IF{\textit{VSup($\widehat{\textit{PF}}$)} $\geq$ \textit{minSup}$\times |\mathcal{D}|$}
		\IF{\textit{qe} == $F$}
		\STATE \textit{new\_match}++;
		\ENDIF
		\STATE \textbf{call} \textbf{DepthFirstSearch}($\widehat{\textit{PF}}$, \textit{qes}, \textit{SF}, \textit{minSup}, \textit{new\_match})
		\ENDIF
		\ENDFOR
\end{algorithmic}
\end{algorithm}

The input of the DepthFirstSearch procedure, as illustrated in Algorithm \ref{DepthFirstSearch}, has a prefix S-TIRP $\widehat{P}$, a query event sequence \textit{qes}, a set containing all frequent S-TIRPs \textit{SF} that can perform extension, the required minimum support \textit{minSup}, as well as the matching position of the query event sequence \textit{match}. The last event of $\widehat{P}$ is represented as \textit{lastEventOfP}, and the current query event is expressed as \textit{qe}. The procedure first determines whether \textit{lastEventOfP}  is equal to \textit{qe}, and if it is, \textit{match} is updated by increasing 1; otherwise, \textit{match} remains 0. Then, TaTIRP evaluates whether the S-TIRP matches all events in \textit{qes} in which \textit{match} is equal to the size of \textit{qes}  and outputs $\widehat{P}$ in case it does.  Provided that the value of vertical support of combining \textit{lastEventOfP} and \textit{qe} is less than \textit{minSup} $\times$ $|\mathcal{D}|$, then the extension operation of the S-TIRP $\widehat{P}$ can be terminated based on the UQPP strategy. Subsequently, for each pattern $\widehat{F}$ in \textit{SF}, it assigns the value of \textit{match} to \textit{new\_match} to prevent missing \textit{match} during the depth-first search process. If the value of PSM(\textit{lastEventOfP}, $F$) is no greater than the threshold, then any pattern extended by these two patterns cannot be frequent following the UEPP strategy. Once all the conditions are fulfilled, we can build the vertical database for the extended pattern \textit{$\widehat{\textit{PF}}$}. In practice, the time-interval between two S-TIRPs may be too long and wide, resulting in less meaningful mining results; thus, we employ \textit{maxGap} and \textit{maxDura} to constrain the scale between time-intervals during construction. Assuming that the vertical support of $\widehat{\textit{PF}}$ meets the frequency condition, then $\widehat{\textit{PF}}$ can be further extended.  Before the next DepthFirstSearch procedure is invoked, it is essential to determine whether \textit{qe} equals $F$; if true, then \textit{new\_match} is updated by adding one. The process continues until all the target patterns have been found.

The time complexity analysis of the TaTIRP algorithm reveals that the primary computational cost arises from constructing the vertical database and performing the depth-first search. Initially, the algorithm scans and filters the database using the USFP strategy, which has a time complexity of O($|\mathcal{D}| \times T_S$), where $|\mathcal{D}|$ is the number of sequences in the database and $T_S$ represents the average number of time-interval events per sequence. Next, the time complexity for constructing the vertical database and the PSM matrix is O($(N_S)' \times (T_S)^2$), where $(N_S)'$ is the number of sequences after filtering. Finally, the time complexity of the depth-first search is O($N_1 × (N_S)' × T_S$), where $N_1$ is the number of candidate patterns after pruning. Through effective pruning strategies such as UQPP and UEPP, the number of candidate patterns is reduced from the original $N$ to $N_1$, significantly lowering the overall computational complexity.

In terms of the correctness and completeness of the proposed algorithm, first, the algorithm TaTIRP is an improvement over the algorithm FastTIRP. In the initial stage, we remove sequences that do not include the query event sequence, because the target patterns necessarily contain every event in the identical chronological order, and sequences that exclude them cannot be the supporting sequences. Thus, the removal does not affect the results. In contrast, we utilize the PSM structure to preserve the co-occurring support counts between two frequent S-TIRPs and use this as a criterion to terminate the pattern extensions in advance. Owing to the downward closure property of support, an infrequent pattern implies that all its extended patterns will also be infrequent. The UQPP strategy is based on the principle that the current query event must come after the pattern is extended. If the combination of these two patterns is invalid, their extended patterns would inevitably be invalid as well. The UEPP strategy, which has already been validated in FastTIRP, states that if the co-occurrence between the current event and the extended pattern is below the required threshold, further extension is unnecessary. Regarding the determination of whether the query event sequence has been matched, this is managed sequentially using a variable \textit{match} to record the times of matches with the query event sequence. It increments by one if it matches the current query event; otherwise, it remains unchanged. When \textit{match} equals the size of the query event sequence, it signifies that a successful match has been achieved. In summary, our algorithm is correct in extracting the complete desired patterns.

\section{Experiment}  \label{Experiment}

This section primarily verifies and analyzes the efficiency of the proposed approach through experiments focusing on runtime and memory consumption. We develop two adaptations of the TaTIRP algorithm to assess the correctness and effectiveness of the designed strategies. First, we adopt FastTIRP \cite{fournier2022fasttirp} as the baseline algorithm and apply a post-processing technique to extract the target patterns corresponding to the query event sequence. The algorithm, composed of baseline and post-processing techniques, is denoted as FastTIRP$^*$. Then, since Strategy \ref{stra_2} is present in both FastTIRP and TaTIRP, its validity has been proven in FastTIRP. Thereafter, in subsequent experiments, we make the default that all variants of TaTIRP contain Strategy \ref{stra_2}. TaTIRP$_1$ indicates that Strategy \ref{stra_1} is utilized and not Strategy \ref{stra_3}, and TaTIRP$_2$ represents the opposite, employing Strategy \ref{stra_3} without Strategy \ref{stra_1}. TaTIRP$_{12}$ applies all the strategies simultaneously. In summary, there are altogether five algorithms for comparison. All algorithms are implemented in Java. All experiments are executed on a computer with a 64-bit Windows 10 operating system, 16 GB of RAM, and a 12th-generation Core i7-12700 processor. 

\subsection{Datasets}

In the experiments, we use six datasets, including four real datasets (ASL, Hepatitis, Diabetes, and Smarthome) and two synthetic datasets (DS1 and DS2). Different datasets have different characteristics, so the algorithm can be evaluated from various perspectives.

\begin{table*}[h]
  \centering
  \caption{Characteristics of the datasets}
  \label{table:Characteristics}
	\begin{tabular}{|c|c|c|c|c|}
		\hline
		\textbf{Dataset} & \textbf{$\#$ of sequences} & \textbf{$\#$ of time-intervals} & \textbf{$\#$ of event} &  \textbf{$\#$ of promising sequences} \\ \hline
		ASL        & 65     & 2,037  & 146 & 50   \\ \hline
		Hepatitis  & 498    & 48,029 & 63  & 373  \\ \hline
		Diabetes   & 2,038   & 80,538 & 35  & 1,923 \\ \hline
        Smarthome  & 89     & 23,213 & 95  & 60   \\ \hline
        DS1        & 1,000   & 20,000 & 20  & 30   \\ \hline
        DS2        & 100,000 & 1,000,000    & 10  & 9,728 \\ \hline
	\end{tabular}
\end{table*}

\begin{itemize}
    \item The ASL (American Sign Language) dataset \cite{papapetrou2009mining} contains a large collection of videos showcasing various gestures and movements in ASL. These videos are usually performed by native ASL signers to capture authentic sign language expressions. Annotations indicate the start and end times of each sign, enabling analysis of fluency and temporal aspects of sign language.

   \item The Hepatitis dataset \cite{patel2008mining} is typically used in medical and health research, primarily for analyzing data related to hepatitis. It consists of records of treatments received by patients and treatment outcomes, patient survival status, survival time, occurrence of complications, prognostic assessments, etc.

   \item The Diabetes dataset \cite{moskovitch2015outcomes} is primarily for analyzing data related to diabetes.  It records treatment plans and treatment outcomes for patients, including medication, insulin injections, blood glucose monitoring records, etc.

   \item The Smarthome dataset \cite{jakkula2011detecting} is characterized by its comprehensive collection of data about home automation systems. It encompasses various aspects such as sensor data, user interactions, environmental conditions, and device states within a smart home environment. 

   \item DS1 \cite{lai2024mining} is a synthetic interval-based database generated by the FTDPMiner-EP algorithm. It has 1,000 sequences where each sequence contains 20 symbolic time-intervals, and there is a total of 100 event types. 

   \item DS2 \cite{lai2024mining} is also a synthetic time-interval sequence dataset of 100,000 sequences in which each sequence includes 10 symbolic time-intervals, and there is a total of 100 event types.
\end{itemize}

Table \ref{table:Characteristics} demonstrates the number of sequences, time-intervals, and events included in each dataset. It is noted that $\#$ stands for the number. For each dataset, we select a query event sequence of size 2 except for DS2, whose query size is 1. This is because each sequence in DS2 holds at most ten time-intervals, rendering it hard to obtain longer patterns. The randomly selected \textit{qes} of ASL is  \{53, 191\}, Hepatitis is \{3, 8\}, Diabetes is \{35, 34\}, Smarthome is \{9, 13\}, DS1 is \{93, 75\}, and  DS2 is \{39\}. For all experiments, we configure $\epsilon$ as 0, \textit{minGap} as 0, \textit{maxGap} as 30, \textit{minDura} as 0, and \textit{maxDura} as 2000. Besides, due to the high density of the Smarthome dataset and the time-intervals within each sequence being numerous, the algorithm encounters memory leakage issues during execution. Therefore, we set the maximum pattern extension length for Smarthome to 5.

\subsection{Runtime analysis}

\begin{figure*}[h]
    \centering
    \includegraphics[trim=160 0 0 0,clip,scale=0.41]{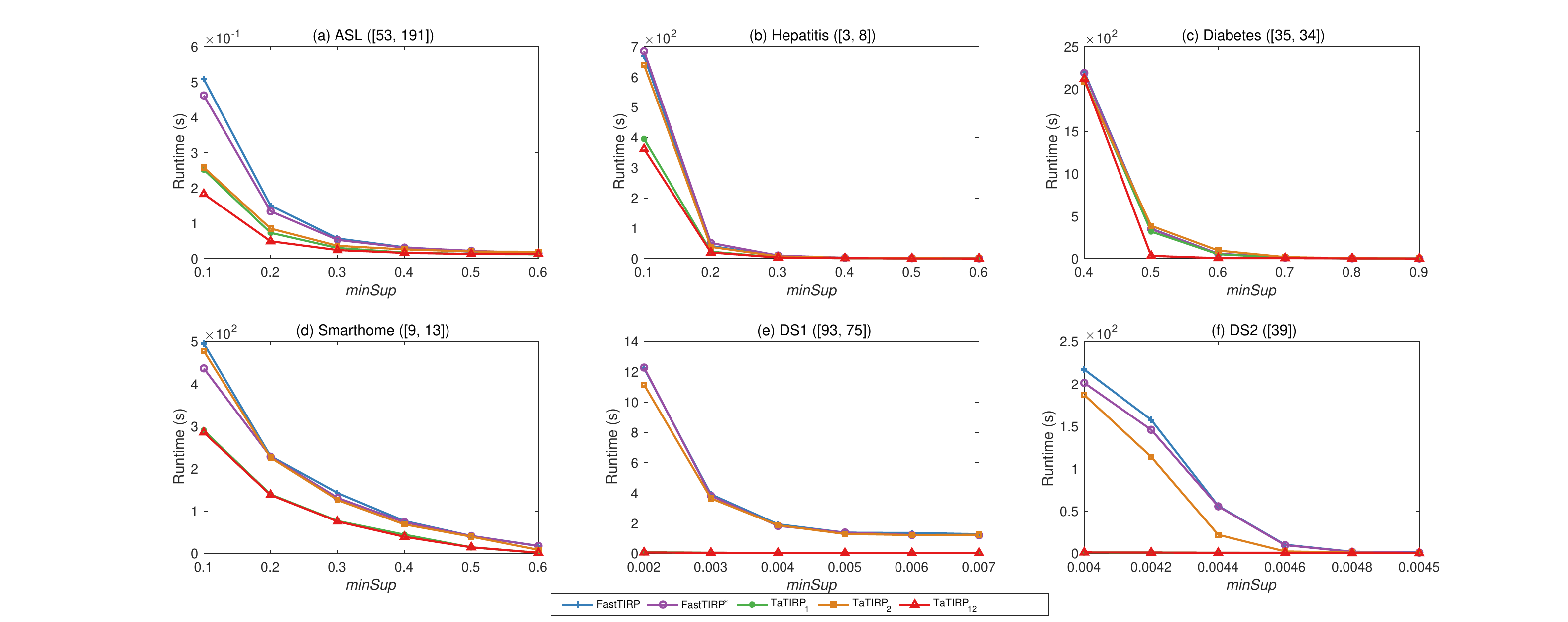}
    \caption{Results of runtime under various \textit{minSup}.}
    \label{fig:TaTIRP_runtime}      
\end{figure*}

We conduct experiments on the six datasets by adjusting \textit{minSup} to evaluate the efficiency of the proposed algorithm. The experiment results are displayed in Fig. \ref{fig:TaTIRP_runtime}.

In general, the runtime of the five comparison algorithms gradually decreases for all datasets with progressively increasing \textit{minSup}. We can observe that FastTIRP takes the longest runtime on all the datasets, even though FastTIRP$^*$ includes a post-processing operation on top of it. This is because the former outputs a larger number of patterns compared to the latter, and the transmission of the output consumes more time than determining whether it is a target TIRP or not. From the figure, it can be noticed that the runtime of TaTIRP$_1$ is shorter than that of TaTIRP$_2$. It is because TaTIRP$_1$ removes some time-interval sequences at the initial stage of the algorithms, which not only allows for the early elimination of some frequent patterns that are irrelevant to the query event sequence and the construction of vertical databases for them, but also minimizes the time spent exploring unpromising sequences during the extension of the remaining frequent patterns. TaTIRP$_1$ takes a holistic perspective of the query sequence event to check the availability of a target pattern, while TaTIRP$_2$ takes a local aspect during pattern extension to determine whether the combination of the current extended pattern and the current query event is frequent and can be part of a target pattern. For example, runtime comparisons between different algorithms are evident in the ASL dataset. FastTIRP and FastTIRP$^*$ exhibit the longest and most comparable runtime, followed by TaTIRP$_1$ and TaTIRP$_2$ with slightly shorter durations, while TaTIRP$_{12}$ demonstrates the shortest runtime, implying the performance of the proposed algorithm can be optimized when all the strategies are applied. Notice that some runtime curves do not exhibit a linear decline. This is because when the \textit{minSup} is low, changes in \textit{minSup} do not substantially affect the number of derived target time-interval patterns. In dense datasets where a sequence encompasses numerous time-intervals, the runtime of TaTIRP$_1$ and TaTIRP$_{12}$ is nearly identical and appears almost superimposed on the figure. For instance, in the Hepatitis and DS1 datasets, the red and green curves converge almost completely, owing to the predominant influence of Strategy \ref{stra_1}.

\subsection{Memory analysis}

\begin{figure*}[h]
    \centering
    \includegraphics[trim=160 0 0 0,clip,scale=0.41]{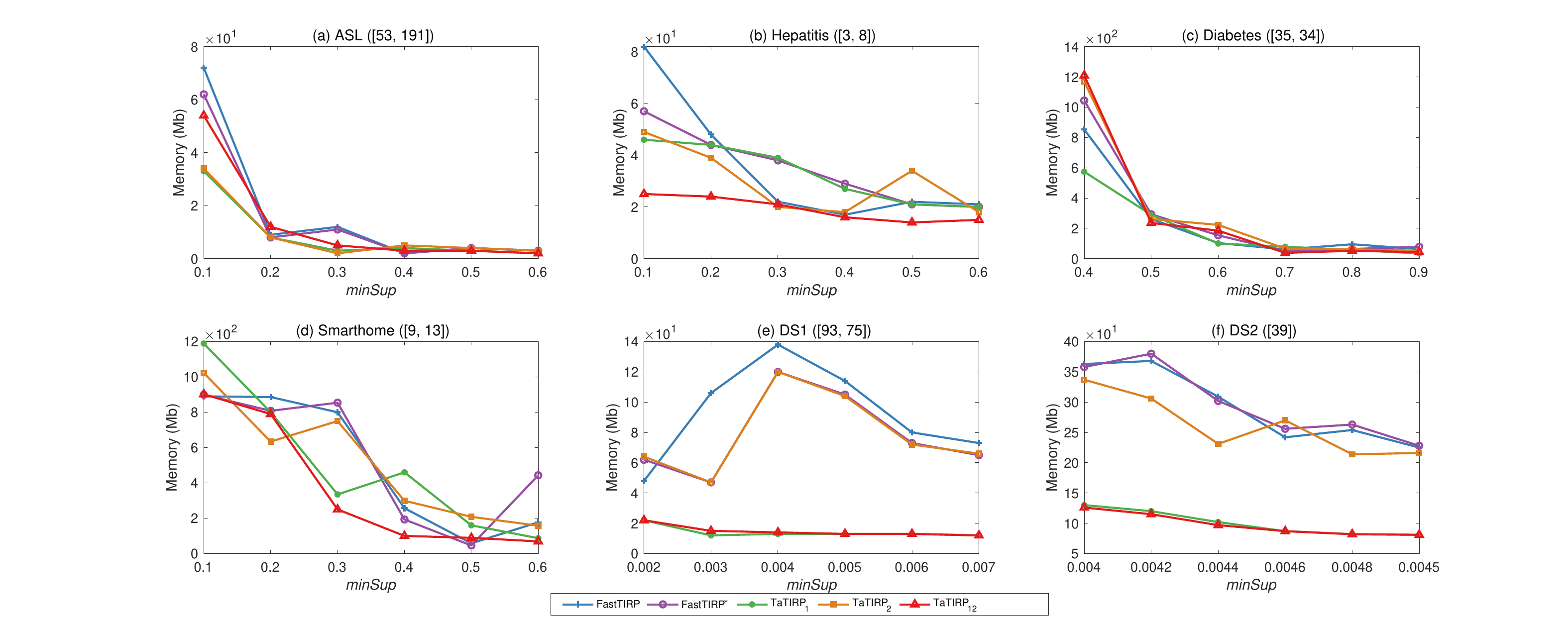}
    \caption{Results of memory consumption under various \textit{minSup}.}
    \label{fig:TaTIRP_Join_count}      
\end{figure*}

In this section, we discuss the comparative performance of different algorithms across six datasets as the minimum support (\textit{minSup}) changes in terms of memory occupation, and the experiment result is exhibited in Fig. \ref{fig:TaTIRP_Join_count}.

Overall, TaTIRP$_{12}$, which integrates all three strategies, stands out in terms of memory optimization, particularly when handling large, complex datasets. It consistently outperforms the other algorithms, especially for high-density and large-scale datasets like Smarthome and DS2. While FastTIRP and FastTIRP$^*$ perform well on simpler datasets, their memory consumption and efficiency have considerable room for improvement, especially when applied to more complex data. FastTIRP$^*$, with its additional post-processing step, consistently uses more memory than FastTIRP, making it less efficient in comparison to TaTIRP$_{12}$. TaTIRP$_{1}$ and TaTIRP$_{2}$, which focus on individual strategies, also provide valuable insights. Regarding TaTIRP$_{2}$, while it utilizes Strategy \ref{stra_3}, the memory consumption compared to TaTIRP$_{1}$ is not significant across all datasets,  particularly for simpler datasets like Diabetes and DS1, where the difference in memory usage between the two algorithms is minimal. This is because these datasets are less complex, with fewer event types and time-intervals. However, for more complex datasets like Smarthome and DS2, Strategy \ref{stra_3} significantly reduces memory consumption by eliminating unpromising sequences that do not match the query event sequence. As dataset complexity increases, the advantages of TaTIRP$_{2}$ and TaTIRP$_{12}$ become more apparent, with TaTIRP$_{2}$ benefiting from Strategy \ref{stra_3}'s pruning and TaTIRP$_{12}$ benefitting from the combination of all three strategies. In conclusion, while TaTIRP$_{1}$ and TaTIRP$_{2}$ offer improvements over FastTIRP and FastTIRP$^*$, the integration of all strategies in TaTIRP$_{12}$ delivers the best performance, especially for large, complex datasets.

\subsection{Pattern analysis}

To verify the correctness and completeness of our proposed algorithms, the number of extracted patterns obtained in different datasets has been listed while adjusting \textit{minSup} as shown in Table \ref{table:pattern}. Owing to the differing threshold configurations across various datasets during the experiments, we designate \textit{minSup}$_1$ through \textit{minSup}$_6$ to represent the thresholds that progressively increase.

\begin{table*}[h]
  \centering
  \caption{The number of extracted patterns}
  \label{table:pattern}
	\begin{tabular}{|c|c|c|c|c|c|c|c|}
		\hline
		\textbf{Dataset} & \textbf{Algorithm} & \textit{minSup}$_1$ & \textit{minSup}$_2$ & \textit{minSup}$_3$ & \textit{minSup}$_4$ & \textit{minSup}$_5$ & \textit{minSup}$_6$   \\ \hline
		\multirow{3}{*}{ASL}       & FastTIRP  &  11396  &  1266  & 275   &  113  &  45   & 22  \\ \cline{2-8}
        \multirow{3}{*}{ }         & FastTIRP$^*$   &  783    &  89    & 17    &  7    &  3    &  2  \\ \cline{2-8}
        \multirow{3}{*}{ }         & TaTIRP   &  783    &  89    & 17    &  7    &  3    &  2  \\ \hline
        \multirow{3}{*}{Hepatitis} & FastTIRP &  407518 &  18838 & 3027  & 731   &  251  & 84 \\ \cline{2-8}
        \multirow{3}{*}{ }         & FastTIRP$^*$   &  14726  &  508   & 64    &  14   &   3   &  1 \\ \cline{2-8}
        \multirow{3}{*}{ }         & TaTIRP   &  14726  &  508   & 64    &  14   &   3   &  1 \\ \hline
        \multirow{3}{*}{Diabetes}  & FastTIRP &  2296   &  686   & 232   &  90   &  32   & 11 \\ \cline{2-8}
        \multirow{3}{*}{ }         & FastTIRP$^*$   &  758    &  195   & 50    & 16    &  4    &  1 \\ \cline{2-8}
        \multirow{3}{*}{ }         & TaTIRP   &  758    &  195   & 50    & 16    &  4    &  1 \\ \hline
        \multirow{3}{*}{Smarthome} & FastTIRP &  247728 &  52703 & 17757 & 7500  & 3350  & 1313 \\ \cline{2-8}
        \multirow{3}{*}{ }         & FastTIRP$^*$   &  19994  &  5410  & 2149  & 862   &  204  & 29   \\ \cline{2-8}
        \multirow{3}{*}{ }         & TaTIRP   &  19994  &  5410  & 2149  & 862   &  204  & 29   \\ \hline
        \multirow{3}{*}{DS1}       & FastTIRP &  343500 &  78673 & 24854 & 12662 & 10460 & 10080  \\ \cline{2-8}
        \multirow{3}{*}{ }         & FastTIRP$^*$   &  329    &  79    & 28    & 7     &   2   & 1   \\ \cline{2-8}
        \multirow{3}{*}{ }         & TaTIRP   &  329    &  79    & 28    & 7     &   2   & 1   \\ \hline
        \multirow{3}{*}{DS2}       & FastTIRP &  9705   &  8223  & 4873  & 1666  & 351   & 122 \\ \cline{2-8}
        \multirow{3}{*}{ }         & FastTIRP$^*$   &  193    &  163   & 98    & 33    & 5     &  3  \\ \cline{2-8}
        \multirow{3}{*}{ }         & TaTIRP   &  193    &  163   & 98    & 33    & 5     &  3  \\ \hline
	\end{tabular}
\end{table*}

We compare the number of output patterns among the three algorithms: FastTIRP, FastTIRP$^*$, and TaTIRP. With the progressive rise in the support threshold, the patterns extracted from both FastTIRP and TaTIRP gradually decline and drop fast at some thresholds. For example, 70 percent of the patterns from \textit{minSup}$_1$ to \textit{minSup}$_2$ are removed in the Smarthome dataset. In contrast to the extensively studied FastTIRP, which has received considerable attention, TaTIRP prioritizes the customer's subjective target pattern. The inclusion of FastTIRP in the table aims to emphasize the disparity between the number of complete patterns and target patterns and demonstrate the ability of the algorithm to filter out patterns that are less relevant to users and accurately retrieve patterns of interest. For example, the quantity of TIRP is 27 times that of TaTIRP in the Hepatitis dataset and 1044 times in the DS1 dataset with \textit{minSup}$_1$. This substantial disparity renders it exceedingly challenging to discern query-related patterns within the immense data expanse. A small number of user-related patterns can more effectively assist us in making better decisions. For example, in the Diabetes dataset, the \textit{qes} can be defined as \{blood glucose monitoring, medication usage, insulin injection\}, representing the key events related to diabetes management. This sequence focuses on tracking critical actions such as monitoring blood glucose levels, administering medication, and injecting insulin. These events are crucial for understanding how diabetes patients' treatment is structured and how the timing between these actions impacts their health. Similarly, in the Smarthome dataset, the \textit{qes} can be defined as \{door switches, light control, temperature change\}, which is used to analyze the interactions between home automation devices. This sequence explores how different devices within a smart home collaborate, such as how door switches trigger light control actions or temperature changes. The timing and relationship between these events help reveal patterns of device behavior and their coordinated functions within the home.

\subsection{Scalability analysis}

\begin{figure}[h]
    \centering
    \includegraphics[trim=80 0 0 0,clip,scale=0.27] {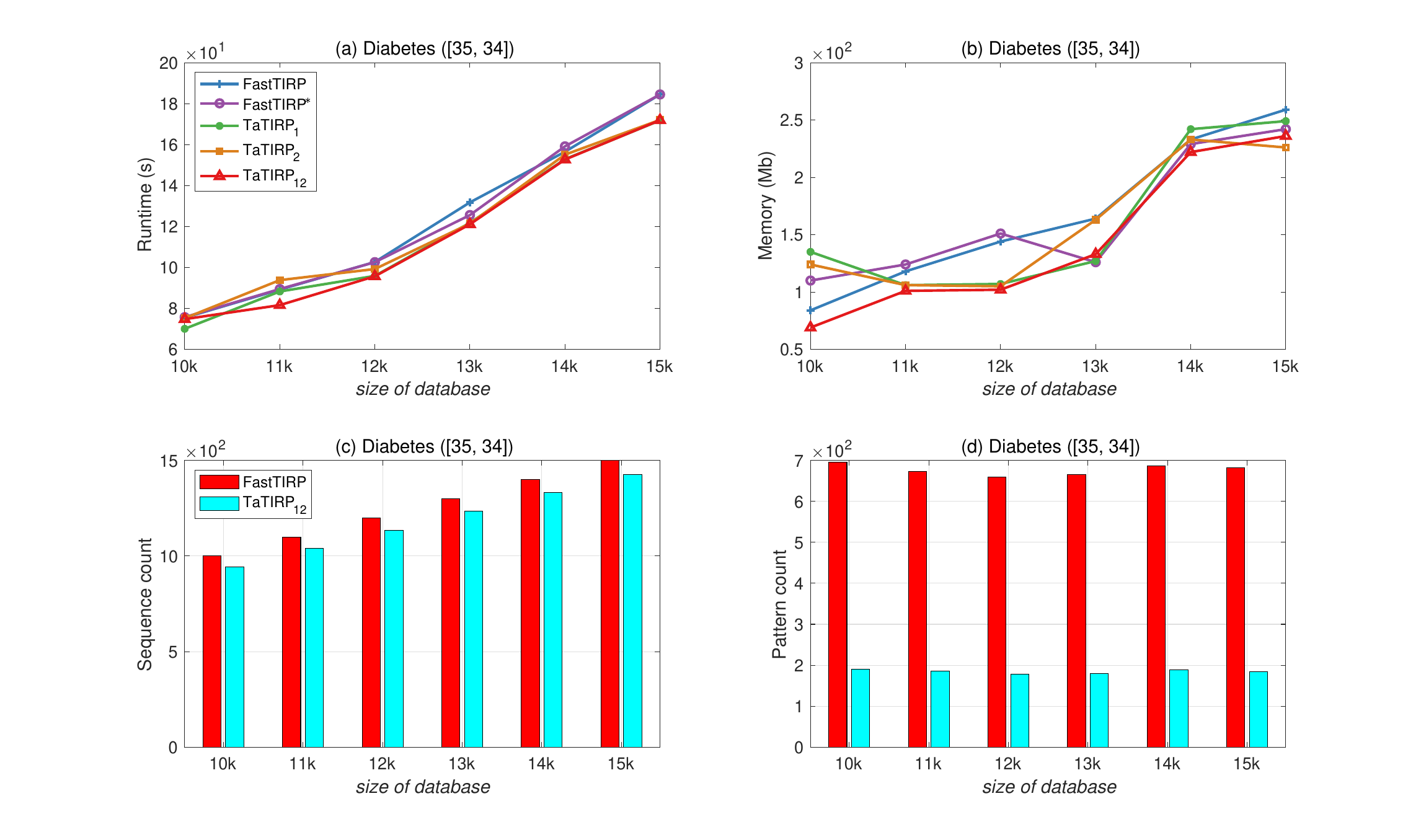}
    \caption{Results of scalability under various \textit{minSup}.}
    \label{fig:TaTIRP_scalability} 
\end{figure}

In this subsection, we evaluate the scalability of the proposed algorithm using varying sizes of the Diabetes dataset, ranging from 10k to 15k, as shown in Fig. \ref{fig:TaTIRP_scalability}. The Diabetes dataset is selected for its extensive collection of thousands of sequences, each containing a significant number of time-intervals, allowing for a comprehensive analysis of the proposed algorithm.

In the experimental configuration, \textit{minSup} is set to 0.5, and \textit{qes} is designated as \{35, 34\}. We compare the runtime, memory consumption, number of sequences, and number of output patterns for each algorithm as the dataset size increases. It is evident that the runtime increases linearly with the dataset size, with slight variations, and our proposed TaTIRP$_{12}$ consistently performs the best, maintaining the lowest runtime. Regarding memory usage, we observe that the memory consumption of FastTIRP is consistently higher than that of TaTIRP$_1$, TaTIRP$_2$, and TaTIRP$_{12}$ across all dataset sizes. Among the proposed methods, TaTIRP$_{12}$ consistently demonstrates the most efficient memory usage, consuming the least memory across all dataset sizes. In comparison, FastTIRP consumes more memory than the other methods as the dataset size increases. However, for smaller datasets (e.g., 10k), FastTIRP exhibits relatively lower memory consumption. As the dataset size grows, particularly for larger datasets (e.g., 14k and 15k), its memory consumption rises sharply. This aligns with the memory analysis results from earlier and highlights the scalability of the proposed algorithm. Fig. \ref{fig:TaTIRP_scalability}(c) further illustrates the size of the original dataset and the corresponding promising sequences after employing Strategy \ref{stra_1} by adjusting the size of the dataset. As the dataset expands, the quantity of promising sequences also rises gradually, which validates the effectiveness of Strategy \ref{stra_1} across datasets of different sizes. Fig. \ref{fig:TaTIRP_scalability}(d) presents the number of patterns excavated in the different-sized datasets, and we remark that the target query acquires far fewer patterns than the complete pattern excavation, contributing to a close approximation of the expectation. For example, the number of derived TIRPs in any dataset size is approximately three times that of TaTIRPs. In summary, the proposed algorithm exhibits robust scalability.

\section{Conclusion} \label{Conclusion}

In this paper, a novel algorithm is proposed for target query time-interval-related patterns. To further enhance the algorithm’s efficiency, three advanced pruning strategies are designed to eliminate unpromising sequences and extension nodes in advance. Among them, the USFP Strategy determines that a sequence is promising depending on the occurrence of the query event sequence. The UQPP and UEPP Strategy then exploit the PSM structure proposed by FastTIRP to interrupt unpromising extensions from the extended pattern and the current query event perspective. These strategies mutually complement each other to render the presented algorithm efficient in terms of runtime and memory. Ultimately, the experiments have demonstrated the superior performance of our proposed algorithm. Despite its promising performance, TaTIRP has several limitations that we aim to address in future work. One key limitation is the scalability of the pruning process, particularly when dealing with large or dynamic datasets. To tackle this issue, we plan to explore additional pruning techniques that will improve the algorithm’s efficiency and enable it to scale more effectively for large-scale and real-time applications. Another limitation is the algorithm's reliance on a predefined query event sequence, which reduces its flexibility, especially when handling multiple queries. To improve this, we intend to expand the querying approach, allowing TaTIRP to handle multi-query time-interval pattern mining. This will enhance the algorithm's ability to manage more diverse and complex data patterns.

\bibliographystyle{IEEEtran}
\bibliography{TaTIRP}

\end{document}